\documentclass[usenatbib]{mn2e}

\usepackage{natbib}
\usepackage{ae}
\usepackage{lmodern}
\usepackage{delarray}
\usepackage{color}
\usepackage{amsmath}
\usepackage{amssymb}
\usepackage{here}
\usepackage{graphicx}
\usepackage[utf8]{inputenc}
\usepackage{float}
\usepackage{epsfig}
\usepackage{tensor}
\usepackage{apjfonts}
\usepackage{ctable}
\usepackage{fixltx2e} 

\newcommand{\ord}{\mathcal{O}}
\newcommand{\mach}{\mathcal{M}}
\newcommand{\xvect}{\mathbf{x}}
\newcommand{\be}{\begin{equation}} \newcommand{\ee}{\end{equation}}

\newcommand{\solarmass}{\mathrm{M}_{\sun}}
\newcommand{\lang}{\left\langle} \newcommand{\rang}{\right\rangle}
\newcommand{\pc}{\mathrm{pc}}

\newcommand{\acknowledgments}{\begin{small}\section*{Acknowledgments}\end{small}}
\newcommand\altaffilmark[1]{$^{#1}$}
\newcommand\altaffiltext[1]{$^{#1}$}

\title{Mapping the core mass function to the initial mass function}

\author[Guszejnov \&\ Hopkins]{
\parbox[t]{\textwidth}{ D\'avid Guszejnov\altaffilmark{1}\thanks{E-mail:guszejnov@caltech.edu} and Philip F. Hopkins\altaffilmark{1}}
\vspace*{6pt} \\
\altaffiltext{1}{TAPIR, Mailcode 350-17, California Institute of Technology, Pasadena, CA 91125, USA} \\
}

\date{Accepted by MNRAS, 17 April 2015 \vspace{-0.6cm}}
\begin{document}
\maketitle
\label{firstpage}

\begin{abstract}

It has been shown that fragmentation within self-gravitating, turbulent molecular clouds (``turbulent fragmentation'') can naturally explain the observed properties of protostellar cores, including the core mass function (CMF). Here, we extend recently-developed analytic models for turbulent fragmentation to follow the time-dependent hierarchical fragmentation of self-gravitating cores, until they reach effectively infinite density (and form stars). We show that turbulent fragmentation robustly predicts two key features of the IMF. First, a high-mass power-law scaling very close to the Salpeter slope, which is a generic consequence of the scale-free nature of turbulence and self-gravity. We predict the IMF slope (-2.3) is slightly steeper then the CMF slope (-2.1), owing to the slower collapse and easier fragmentation of large cores. Second, a turnover mass, which is set by a combination of the CMF turnover mass (a couple solar masses, determined by the `sonic scale' of galactic turbulence, and so weakly dependent on galaxy properties), and the equation of state (EOS). A ``soft'' EOS with polytropic index $\gamma<1.0$ predicts that the IMF slope becomes ``shallow'' below the sonic scale, but fails to produce the full turnover observed. An EOS which becomes ``stiff'' at sufficiently low surface densities $\Sigma_{\rm gas} \sim 5000\,M_{\odot}\,{\rm pc^{-2}}$, and/or models where each collapsing core is able to heat and effectively stiffen the EOS of a modest mass ($\sim 0.02\,M_{\odot}$) of surrounding gas, are able to reproduce the observed turnover. Such features are likely a consequence of more detailed chemistry and radiative feedback.

\end{abstract}

\begin{keywords}
star formation: general --- galaxies: formation --- galaxies: evolution --- galaxies: active --- 
cosmology: theory
\vspace{-1.0cm}
\end{keywords}

\section{Introduction}\label{sec:intro}

The mass distribution of newly formed stars, often referred to as the \textit{Initial Mass Function} or \textit{IMF}, is fundamental in many aspects of astrophysics. Understanding the processes leading to the observed IMF provides valuable insight into not only star formation but into the evolution of galactic structures and the formation of planets. So far observations of different galaxies and regions within the Milky Way suggest that some qualitative features of the IMF are universal (\citealt{IMF_universality}, \citealt{IMF_review}). These include:
\begin{itemize}
	\item a power law-like slope ($dn/dM\propto M^{-2.3}$) for large masses;
	\item turnover around 0.1-1.0 solar mass;
	\item lognormal-like or power law-like behavior for small masses.
\end{itemize}
The universality of these properties implies that some fundamental physical process influences the initial stellar mass distribution. It is important to note that, of these three properties, the power law-like slope is also ubiquitous to wildly different systems including dark matter halos (\citealt{PressSchechter}), giant molecular clouds (\citealt{GMC_Rosolowsky}), young star clusters (\citealt{Portegies_young_clusters}) and HI holes in the interstellar medium (\citealt{Weisz_hole_ISM}). The exponent of $dn/dM\propto M^{-2.3}$ is close to that which implies that an equal amount of mass is distributed in every logarithmic interval in mass, which points to a self-similar process being the main driving force behind these distributions.

\begin{figure}
\begin {center}
\includegraphics[width=0.95 \linewidth]{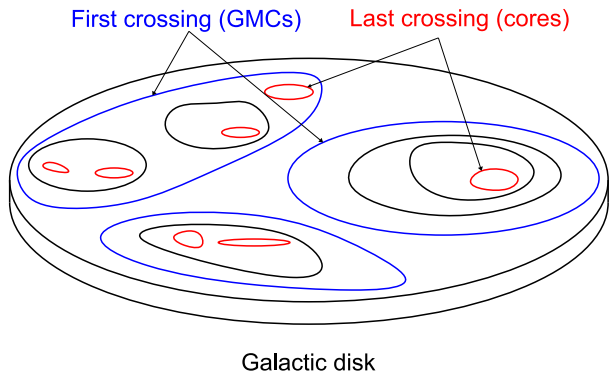}
\caption{Cartoon illustration of hierarchical turbulent fragmentation in a galactic disk. The scale of the largest self gravitating clouds is called the ``first crossing'' (largest scale where the density $\rho>\rho_{\rm crit}$, see Eq. \ref{eq:gen_collapse_threshold}), which corresponds to giant molecular clouds (GMCs) while the scale of the smallest clouds (usually embedded in larger ones) is the last ``crossing scale'' which correspond to protostellar cores. }\label{fig:disk_crossing}
\end {center}
\end{figure}

A candidate for such process is turbulent fragmentation. It is widely accepted that stars are formed by the gravitational collapse of dense molecular clouds (\citealt{McKee_star_formation}). Gas in these clouds is highly turbulent which leads to large fluctuations in density that in turn then lead to the emergence of subregions that are independently collapsing (see Fig. \ref{fig:disk_crossing}). Denser regions collapse faster, turning into stars whose feedback (e.g. radiation, solar winds) heats up or blows the surrounding gas away effectively preventing further star formation in that area. 

This process is inherently hierarchical, which suggests that it should be possible to derive a single model which simultaneously links the largest scales of collapse all the way down to the smallest (the scales of individual stars). This is not possible in simulations because of resolution limitations, but can be approximately treated in analytic models. 

This paradigm was explored by \cite{Padoan_theory} and \cite{Padoan_Nordlund_2002_IMF}, then made more rigorous by \cite{HC08} who attempted to approximate the IMF in a manner analogous to \cite{PressSchechter}. \cite{excursion_set_ism} expanded upon these works by using an excursion set formalism to calculate the distribution of first crossing mass scales in galactic disks\footnote{In the usual terminology the largest collapsing scale is referred to as the scale of \textit{first crossing} while the smallest collapsing subregion is at the scale of \textit{last crossing}.}. This yielded mass functions very similar to the mass distribution of giant molecular clouds (\textit{GMC}s) which are the largest known bound collections of gas in a galaxy. Meanwhile \cite{core_IMF} found that the mass function of structures at the last crossing scale show a striking similarity to the distribution of protostellar cores (also referred to as \textit{cores}). This core mass function (\textit{CMF}) is remarkably similar to the IMF, the only difference being the position of the turnaround which is at a mass scale 3 time larger than the case of the IMF (\citealt{Sadavoy_observed_CMF, Rathborne_CMF, Alves_CMF_IMF_obs}). Building on these results \cite{general_turbulent_fragment} generalized the formalism to be applicable to a wide range of phenomena by incorporating gases with arbitrary equation of state, magnetic fields, intermittency etc. They also showed that this naturally predicts observed cloud and protostellar core properties such as the ``Larson's laws'' scalings of cloud size, mass, and linewidth (\citealt{Larson_law, Enoch_CMF, Brunt_turb_driving_obs}), stellar clustering and correlation functions from scales $\sim 0.1-1000$\,pc (\citealt{clustering_Lada, Portegies_young_clusters}) as a consequence of turbulent fragmentation.

Nevertheless a major shortcoming of these models is that they only extend to the CMF\footnote{Other attempts were made to connect the IMF and CMF, notable examples are \protect\cite{Padoan_Nordlund_2011_IMF} which used the IMF predicted by \protect\cite{Padoan_Nordlund_2002_IMF} to 'guess' the CMF, and \cite{Clark_clump_lifetime_IMF} which discussed some general properties of the mapping. Both drew attention to the problem of time dependence as the time scales of forming stars of different sizes differs greatly (this has been shown to be important by the simulations of \citealt{Padoan_luminosity}). Our model attempts to partially address this issue.}. It is by no means clear that the ``mapping'' from CMF to IMF is simple or universal. And in fact some of the simple assumptions in these previous works -- for example, that of isothermal gas -- must break down on small scales. Therefore, in this paper we expand upon these works and we argue that it is possible to bridge the gap between the CMF and the IMF by analytically following the collapse of protostellar cores. Gravitational collapse takes place during a finite amount of time during which collapse pumps energy into turbulence causing the cloud to fragment. We are able to build a simple model meant to capture this, and from it derive the principal qualitative features of the IMF. We will show that the high mass IMF slope can be explained purely by turbulent fragmentation and the turnover position is dependent on the underlying thermodynamics and galactic properties, while the low mass end is highly influenced by the aforementioned processes and feedback physics.

The paper is organized as follows. A general overview of the excursion set formalism is given in Sec. \ref{sec:methodology} including several further assumptions regarding the collapsing medium (Sec. \ref{sec:eq_state}) and the time evolution of collapsing protostellar cores (Sec. \ref{sec:time_evol_collapse}). In Sec. \ref{sec:mapping} the model we developed for mapping between CMF and IMF is described in detail. The final results and their implications are discussed in Sec. \ref{sec:conclusions}.

\section{Methodology}\label{sec:methodology}

To map the CMF to the IMF one needs to describe the transition from protostellar cores into protostars. To do that we employ the excursion set formalism outlined in \cite{excursion_set_ism} and \cite{core_IMF} with the addition of time dependence from \cite{general_turbulent_fragment}. Only a broad summary of the method will be given here, see the references for more details.

	\subsection{Density Field Evolution}\label{sec:dens_evol}
	
The aim of the model is to describe the properties of self gravitating turbulent medium (see Sec. 2 of \cite{general_turbulent_fragment} for detailed description). In the case of an isothermal medium, ignoring (for now) self-gravity, the density fluctuations in both sub and supersonic cases have lognormal statistics\footnote{As shown in \cite{Hopkins_isothermal_turb} the statistics are not perfectly lognormal even in the isothermal case, however those particular corrections have very little effect on our results.} which means that the density contrast $\delta(\xvect)=\ln\left[\rho(\xvect)/\rho_0\right]+S/2$, where $\rho(\xvect)$ is the local density, $\rho_0$ is the mean density and $S$ is the variance of $\ln\rho$, would follow a normal distribution, thus
\be
P(\delta|S)=\frac{1}{2\pi S}\exp{\left(-\frac{\delta^2}{2S}\right)}.
\ee
It is a property of Gaussian and lognormal random variables that an integral over such fields is also Gaussian/lognormal. Thus let us define the average density on scale $\lambda$ as
\be
\rho(\lambda,\xvect)=\int{\rho(\mathbf{x'}) W_{\lambda}(\mathbf{x'}-\xvect)d^3x'},
\ee
where $W_{\lambda}(\mathbf{x'}-\xvect)$ is the window function for averaging. Then, according to the theorem $\delta(\lambda,\xvect)$ will be also Gaussian. For the sake of brevity from this point on let us drop the $\xvect$ coordinate from these quantities. Also, to simplify the formulas the Fourier transform of the window function ($W\left(\mathbf{k}\right)$) is assumed to be a Heaviside function (cutoff at $k$) \footnote{The calculation could be repeated with $W(k)$ corresponding to real space spheres or filaments but that would have $<10\%$ effect on the final results.}.

Instead of dealing with $\delta$ directly it is more convenient to introduce a new quantity $\Delta\delta\left(\lambda_2|\delta\left[\lambda_1\right]\right)=\delta(\lambda_2)-\delta(\lambda_1)$ which is the contribution to the logarithmic density by scales between $\lambda_1$ and $\lambda_2$. This way we can express $\delta$ as
\be
\label{eq:delta_sum}
\delta(\lambda_i)=\sum_j^{\lambda_j>\lambda_i}{\Delta\delta_j},
\ee
where we use the fact that the density on the largest scale is by the definition the mean density with no variance thus $\delta(\lambda_{\rm{max}})=0$.

In a turbulent system the variance of the logarithmic density field ($\sigma^2(\lambda)$) will tend to an equilibrium value $S(\lambda)$ prescribed by the turbulence. It is well known in the isothermal case that the variance of density is related to the variance of velocity as $S\approx \ln\left(1+\mach^2_{\rm{compressive}}\right)$ where $\mach^2_{\rm{compressive}}$ is the compressive Mach number related to the turbulent velocity dispersion (\citealt{Federrath_turbulence_compressive_PDF}). Following the derivation of \cite{general_turbulent_fragment}:
\be
S(\lambda)=\int_{0}^{\lambda}{\Delta S(\hat{\lambda})d\ln{\hat{\lambda}}}\approx\int_{0}^{\lambda}{\ln{\left[1+\frac{b^2 v_t^2\left(\hat{\lambda}\right)}{c_s^2+\kappa^2\hat{\lambda}^2}\right]}d\ln{\hat{\lambda}}},
\label{eq:S_def}
\ee
where $v_t\left(\lambda\right)$ is the turbulent velocity dispersion on scale $\lambda$, $c_s$ is the thermal sound speed, $b$ is the fraction of the turbulent velocity in compressive motions, which we take to be about $1/2$ (appropriate for randomly driven, super-sonic turbulence, though we have experimented with $b\sim 1/4-1$ and find it makes no qualitative difference to our conclusions), and $\kappa$ is the epicyclic frequency which represents angular momentum suppressing large-scale density fluctuations. Note that this particular scaling for $S(\lambda)$,  as well as the functional form for the density statistics on different scales $\rho(\lambda)$ which we adopt, have been directly measured in numerical simulations (\citealt{Kowal_turb_sim, Federrath_sim_2010});

Let us suppose that instead of an isothermal medium we have gas which follows a polytropic equation of state as
\be
\label{eq:cs_polytropic}
c_s^2=c_{s0}^2\left(\frac{\rho}{\rho_0}\right)^{\gamma-1},
\ee
where $c_{s0}$ is the sound speed at the mean density ($\rho_0$) and $\gamma$ is the polytropic index. In this case (for $0.3<\gamma<1.7$), the statistics can still be approximated as {\em locally} lognormal (i.e.\ lognormal for differentially small perturbations; \citealt{Passot_polytropic_lognormal}), if we apply the replacement $c_s^2\rightarrow c_{s0}^2\left(\rho/\rho_0\right)^{-\left(\gamma-1\right)}$ to Eq. \ref{eq:S_def}. which means that we get $S(\lambda)\rightarrow S(\lambda,\rho)$ so $S$ becomes a functional of $\rho$. This scheme is also an acceptable approximation for gases with more complex equation of states (e.g. $\gamma(\rho)$). Note that this means the total PDF can differ significantly from a lognormal; for $\gamma>1$ large positive-density fluctuations become rarer while $\gamma<1$ makes them more common (producing a power-law high-density tail).\footnote{These effects and the validity of our analytic expressions have been directly verified in simulations (\citealt{Scalo_gamma_sim}, Lynn \& Quataert, private communication).} It should be noted that previous treatments (e.g. \citealt{HC08}) ignored the effect of $\gamma$ on the distribution of $\rho$ despite the fact that it can produce radically different PDFs. For more details see Sec. 3 of \cite{general_turbulent_fragment}.

	\subsection{The Collapse Threshold}
	
	Various authors (e.g.\ \citealt{Chandrasekhar_grav_collapse}, \citealt{Elmegreen_collapse_Toomre}) have shown that including the effects of turbulence and finite vertical disk thickness into a Toomre-type analysis yields a simple scaling for the critical density ($\rho_{\rm crit}$) above which a spherical subregion of size $\lambda$ embedded in a larger disk or cloud becomes gravitationally unstable and collapses. This can be written  
	\be
	\label{eq:gen_collapse_threshold}
	\frac{\rho_{\rm{crit}}(\lambda)}{\rho_0}=\frac{Q}{2 \tilde{\kappa}}\left(1+\frac{h}{\lambda}\right)\left[\frac{\sigma_g^2(\lambda)}{\sigma_g^2(h)}\frac{h}{\lambda}+\tilde{\kappa}^2\frac{\lambda}{h}\right],
	\ee
	where $h$ is the vertical scale of the disk, $\sigma_g^2(\lambda)\approx c_s^2+v_t^2(\lambda)$ is the total velocity dispersion on scale $\lambda$ where $v_t^2(\lambda)$ is the turbulent velocity dispersion at that scale, $\tilde{\kappa}=\kappa/\Omega$ where $\Omega=v_{\rm{circ}}/r_{\rm disk}$ is the orbital frequency at the location $r_{\rm disk}$, $\kappa$ is the epicyclic frequency, and $Q=\sigma_g(h)\kappa/\left(\pi G \Sigma\right)$ is the Toomre parameter, where $\Sigma$ is the surface density of the disk. For the scales of interest here, $\lambda$ is in the inertial-range of turbulence where turbulent kinetic energy scales as $E(\lambda)\propto \lambda^{p}$ with $p$ being the turbulent spectra index; generally $p\in [5/3;2]$, but in this paper we assume $p=2$ for our calculations based on the observed linewidth-size relations (\citealt{Larson_law}; \citealt{Bolatto_2008}; \citealt{Enoch_CMF}), theoretical expectations \citep{Murray_supersonic_Burgers_analytic, Burgers_old, Burgers_book}, and numerical simulations \citep{Schmidt_supersonic_sim}). This leads to the following scaling of the turbulent velocity dispersion and Mach number $\mach$
	\be
	\label{eq:mach_scaling}
	\mach^2(\lambda)\equiv\frac{v_t^2(\lambda)}{\lang c_s^2\left(\rho_0\right)\rang}=\mach^2(h)\left(\frac{\lambda}{h}\right)^{p-1}.
	\ee
	Since we are only interested in protostellar cores, which are much smaller than their parent galactic disk, it is justified to take the limit of $\lambda\ll h$ leading to
	\be
\frac{\rho_{\rm{crit}}(\lambda)}{\rho_0}=\frac{Q'}{1+\mach_{\rm{edge}}^2}\tilde{\lambda}^{-2}\left[\left(\frac{T(\lambda)}{T_0}\right)+\mach_{\rm{edge}}^2\tilde{\lambda}^{p-1}\right],
\label{eq:collapse_threshold_T}
\ee
where $T(\lambda)$ is the temperature averaged over the scale $\lambda$, while $T_0$ is the mean temperature of the whole collapsing cloud.

If we further assume that the gas has a polytropic equation of state then Eq. \ref{eq:collapse_threshold_T} becomes
\be
\frac{\rho_{\rm{crit}}(\lambda)}{\rho_0}=\frac{Q'}{1+\mach_{\rm{edge}}^2}\tilde{\lambda}^{-2}\left[\left(\frac{\rho_{\rm{crit}}(\lambda)}{\rho_0}\right)^{\gamma-1}+\mach_{\rm{edge}}^2\tilde{\lambda}^{p-1}\right],
\label{eq:collapse_threshold}
\ee
where $\tilde{\lambda}=\lambda/h$ is the normalized size scale, $Q'=Q/(2 \tilde{\kappa})$ and $\mach_{\rm{edge}}=\mach(h)$ is the Mach number for the turbulent velocity dispersion at the largest scale\footnote{Once again we note that direct simulation (\citealt{Federrath_sim_2012, Hennebelle_theory, Zentner_cosmos_excursion}) have confirmed that this is a good approximation for the collapse criterion. Even for highly non-spherical, filamentary clouds, the corrections are of $\ord\left(10\%\right)$ to the final predicted mass function.}. This is an implicit equation in case $\gamma\neq 1$ which always has a unique solution for $\gamma<2$. Note that this equation applies identically for sub-structures {\em inside} a core, where in that case $\rho_{0}$, $Q'$, and $\mach_{\rm edge}$ are defined at the scale of the core. For collapsing cores the core scale itself has to be unstable which prescribes $Q'=1$, which we will adopt for the rest of the paper.

For $\mach_{\rm{edge}}^2\tilde{\lambda}^{p-1}\gg 1$ turbulence dominates over thermal support and the critical density becomes roughly
\be
\label{eq:rho_crit_supersonic}
\rho_{\rm{crit}}(\lambda)\approx \rho_0 \tilde{\lambda}^{p-3},
\ee
while in the opposing, subsonic limit
\be
\label{eq:rho_crit_subsonic}
\rho_{\rm{crit}}(\lambda)\approx \rho_0 \left[\left(1+\mach_{\rm{edge}}^2\right)\tilde{\lambda}^2\right]^{-1/\left(2-\gamma\right)}.
\ee
Since we are in the $\lambda\ll h$ limit, the mass of a structure with size scale $\lambda$ and density $\rho(\lambda)$ is just $M(\lambda)=(4\pi/3)\, \lambda^3\,\rho(\lambda)$. And since protostellar cores begin themselves as ``last-crossings'' (smallest collapsing subregions of the galactic disk) in this formalism, they are at the critical density (if they were above it, some smaller scale would necessarily also be self-gravitating), so we can use this equation with $\rho(\lambda)=\rho_{\rm crit}(\lambda)$ and Eq. \ref{eq:rho_crit_supersonic}-\ref{eq:rho_crit_subsonic} to obtain their size-mass relation (see Sec. \ref{sec:mapping}).

	\subsection{The Equation of State} \label{sec:eq_state}
	
	For the purpose of modeling a collapsing protostellar core, a simple polytropic equation of state is not sufficient due to the highly complex heating and cooling processes involved. As a first approximation one can describe the whole cloud as having an \textit{effective polytropic index} which is dependent on global properties (e.g. size, mass). Since the primary physical quantity for radiation absorption is surface density $\Sigma$, we choose to have a polytropic index dependent on this global quantity. Sufficiently dense clouds become optically thick to their own cooling radiation, meaning that blackbody radiation is the primary cooling mechanism. For realistic temperatures molecular hydrogen has a polytropic index of $\gamma=7/5$. In case of less dense clouds, line cooling is the dominant cooling mechanism whose rate is $\propto n^2$, where $n$ is the cloud's number density, while the dominant heating mechanism is cosmic radiation which depends only linearly on the density. This means that an increase in density leads to an effective decrease in temperature, thus $\gamma<1$. Based on these assumptions and on the works of \cite{Masunaga_EQS_highgamma_ref} and \cite{Glover_EQS_lowgamma_ref}, who calculated effective equation of state using full chemical networks in radiation hydrodynamics simulations, we define a simple interpolating equation of state which reproduces the aforementioned two limits:
	
\be
\label{eq:gamma}
\gamma(\Sigma) =\begin{cases} 0.7 &\Sigma < 3\, \solarmass/\pc^2 \\
															0.094 \log_{10}{\left(\frac{\Sigma}{3\,\solarmass/\pc^2}\right)}+0.7 & 3 < \frac{\Sigma}{\solarmass/\pc^2} < 5000   \\
															1.4 &\Sigma > 5000\, \solarmass/\pc^2 \\
		  \end{cases},
\ee
where $\Sigma=M/(4 \pi R^2)$ is defined for each ``fragment'' (cloud or sub-cloud, if it has collapsed independently). This $\gamma(\Sigma)$ equation of state does capture the physics of the limit where the cloud is optically thick to its own cooling radiation, however in the optically thin limit the local density $\rho$ determines the effective polytropic index, not $\Sigma$. Nevertheless this EOS is still useful as the optically thin limit is populated by massive clouds whose fragmentation is barely dependent on the value of $\gamma$ (see Fig. \ref{fig:different_gamma_large}) so changing to a $\rho$ dependent EOS for less dense clouds would not make a significant difference. In any case the effects of variations in the equation of state are investigated in Sec. \ref{sec:robust:EOS}.

It should be noted that the global parameter of our EOS ($\Sigma$ surface density) changes on the dynamical time scale so for sufficiently small $\Delta t$ time step the temperature field evolution can be approximated with the polytrope
\be
\label{eq:T_evol}
T(\lambda,t+\Delta t)=T(\lambda,t)\left(\frac{\rho(\lambda,t+\Delta t)}{\rho(\lambda,t)}\right)^{\gamma\left(\Sigma\right)-1}.
\ee

\subsection{Time-Dependent Collapse of Cores} \label{sec:time_evol_collapse}
	
\begin{figure*}
\begin {center}
\includegraphics[width=\linewidth]{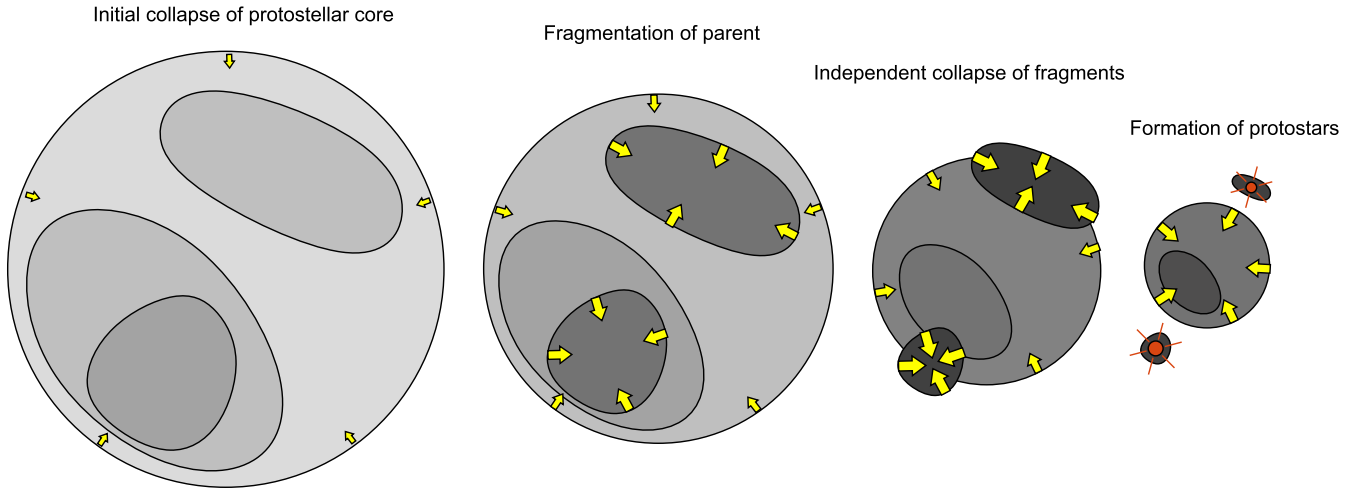}
\caption{Evolution of collapsing protostellar cores, with time increasing from left to right (darker subregions are higher-density, arrows denote regions which are independently self-gravitating and become thicker with increasing collapse rate). As the initial core collapses,  density fluctuations increase (because gravitational energy pumps turbulence), creating self-gravitating subregions. These then collapse independently from the parent cloud, forming protostars at the end.}\label{fig:fragment_fig}
\end {center}
\end{figure*}

	One of the key physical processes in mapping the CMF to the IMF is the nonlinear density field evolution during the collapse phase, which can cause the fragmentation of the cloud (see Fig. \ref{fig:fragment_fig}). To get a handle on this problem, let us first look at the time evolution of the density field in a stationary (statistically time-steady e.g. not globally collapsing/expanding) background. Using the notation of Sec. \ref{sec:dens_evol} and Eq. \ref{eq:delta_sum}, we consider not the density contrast itself, but its modes in Fourier space, as their time evolution simply follows the generalized Fokker-Planck equation (see Sec. 9 of \citealt{general_turbulent_fragment}) 
\be
\label{eq:delta_evol}
\Delta\delta(\tilde{\lambda},t+\Delta t)=\Delta\delta(\tilde{\lambda},t)\left(1-\Delta t/\tau_{\lambda}\right)+\mathcal{R}\sqrt{2 \Delta S(\tilde{\lambda})\Delta t/\tau_{\lambda}},
\ee
where $\mathcal{R}$ is a Gaussian random number with zero mean and unit variance while $\tau_{\lambda}\sim \lambda/v_t(\lambda)$ is the turbulent crossing time on scale $\lambda$, and the turbulence dispersion obeys $v_t^2(\lambda)\propto \lambda$ thus $\tau_{\lambda} \propto \sqrt{\tilde{\lambda}}$ which we normalize as $\tau_{\lambda}(\lambda_{\rm{max}})=1$ thus setting the time units for our problem (see collapse time in Eq. \ref{eq:scale_evol}). This formalism holds for polytropic gases too if we apply the substitution $\Delta S(\tilde{\lambda})\rightarrow \Delta S(\tilde{\lambda},\rho)$ and set it according to Eqs \ref{eq:S_def}-\ref{eq:cs_polytropic} and Eq. \ref{eq:mach_scaling}. For verification of evolution timescale in simulations, see \cite{Pan_supersonic_mixing}.

Note that, as the sub-regions collapse the total ensemble density distribution -- even for isothermal gas -- will deviate significantly from a lognormal. In fact what we predict is that self-gravitating regions develop a power-law tail in their ``total'' (ensemble) density PDFs, as sub-regions collapse on power-law (free-fall) time-scales. This is, of course, exactly what is observed in real dense molecular clouds (see \citealt{Kainulainen_power_law_tail}), and it has been previously shown in simulations that it results naturally from such a fragmentation cascade (see e.g. \citealt{Kritsuk_dens_lognormal, Ballesteros-Paredes_2011, Federrath_sim_compare, Veltchev_2011, Schmalzl_2010}).

	\subsubsection{Turbulent Density Fields in a Collapsing Background}
	
\begin{figure}
\begin {center}
\includegraphics[width=\linewidth]{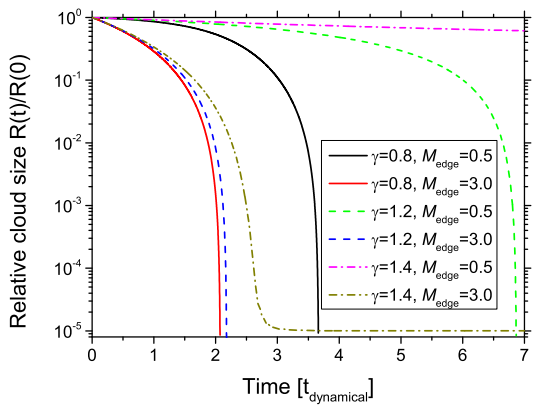}
\caption{Contraction of a self gravitating, collapsing, turbulent parent cloud in time according to Eq. \ref{eq:mach_edge_evol} for different polytropic indices $\gamma$ and edge Mach numbers $\mach_{\rm{edge}}$ (Mach number of the turbulence on the cloud scale). For high Mach numbers the equation of state (e.g. different $\gamma$ values) has little effect on the collapse rate, because the cloud is supported by turbulence. However, for $\gamma>4/3$ the contraction ceases at a finite scale.}\label{fig:cloud_evol}
\end {center}
\end{figure}

In the case of collapsing protostellar cores the density evolution is influenced by the gravitational collapse which pumps energy into turbulence, potentially leading to large density fluctuations and further fragmentation of the cloud. \cite{general_turbulent_fragment} developed a simple model for collapsing spherical clouds which assumes a constant virial parameter (based on \citealt{Robertson_adiabatic_heat_fluid} and \citealt{Murray_star_formation})\footnote{It should be noted that based on current data it is not at all clear that these collapses really happen at constant virial parameter, however we believe it is a reasonable approximation.}. Of course a perfectly spherical collapse would not drive turbulence, but any inhomogeneity in a 'roughly' homogenous media would be greatly amplified by the collapse which will drive the turbulence. Instead of dealing with the microscopic details our model assumes that virial equilibrium is realized between turbulence and gravity on the largest scale, thus the contraction is set by the rate of turbulent energy dissipation whose characteristic time scale is the crossing time $\tau_{\lambda}$. This leads to an equation for the contraction of the cloud:
\be
\frac{d\tilde{r}}{d\tilde{\tau}}=-\tilde{r}^{-1/2}\left(1-\frac{1}{1+\mach_{\rm{edge}}^2}\right)^{3/2},
\label{eq:scale_evol}
\ee
where $\tilde{r}(t)=r(t)/r_0$ is the relative size of the cloud at time $t$ while $\tilde{\tau}\equiv t/t_0$ is time, normalized to the initial cloud dynamical time $t_0\sim 2 Q'^{-3/2}\left(G M_0/R_0^3\right)^{-1/2}$ (see Fig. \ref{fig:cloud_evol} for solutions and \cite{general_turbulent_fragment} for derivation). In this case the initial dynamical time ($t_{0}$) and crossing time only differ by a freely-defined order unity constant, so in our simulations we consider them to be equal without loss of generality. Virial equilibrium implies that that during the collapse of the cloud:
\be
\frac{d \left(\mach_{\rm{edge}}^2\right)}{d\tau}=\left(1+\mach_{\rm{edge}}^2(t=0)\right)\left(-1+3\left(\gamma-1\right)\right)\tilde{r}^{-2+3\left(\gamma-1\right)}\frac{d\tilde{r}}{d\tau},
\label{eq:mach_edge_evol}
\ee
which for constant $\gamma$ simplifies to
\be
1+\mach_{\rm{edge}}^2(t)=\left(1+\mach_{\rm{edge}}^2(t=0)\right)\tilde{r}^{-1+3\left(\gamma-1\right)}.
\label{eq:mach_edge_evol_const_gamma}
\ee

\begin{figure}
\begin {center}
\includegraphics[width=\linewidth]{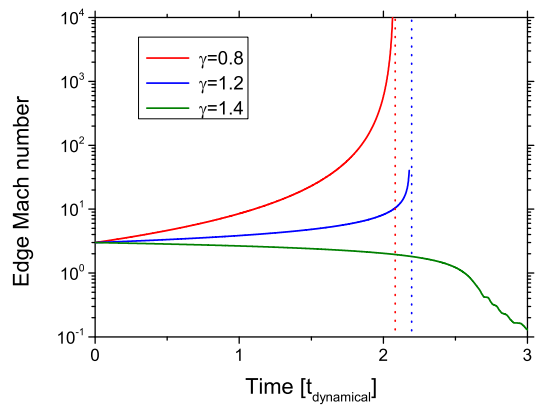}
\caption{Evolution of the edge Mach number (Mach number of turbulence on the cloud scale) in collapsing clouds for different polytropic indices. For $\gamma<4/3$ the contraction of the cloud pumps energy into turbulence, thus the $\mach_{\rm edge}$ diverges as we approach the time of collapse (marked with dotted lines). In the opposite case the sound speed increases faster than the turbulent velocities, pushing the cloud into the subsonic limit (where fragmentation becomes inefficient).}\label{fig:edge_mach}
\end {center}
\end{figure}

In the case $\gamma>4/3$, after some time the sound speed $c_s$ will begin growing faster than $v_t$, stabilizing against collapse. Thus the contraction will seize at a finite $\tilde{r}$ value (see Figs. \ref{fig:cloud_evol}-\ref{fig:edge_mach}). In this case we consider the collapse ``done'' when this size limit is reached. However, if $\gamma<4/3$ then $\tilde{r}=0$ is reached in a finite amount of time. This also means that the cloud can not fragment on arbitrarily small scales as there is not enough time for these fluctuations to grow. For sufficiently small $\tilde{r}$ the collapse becomes scale-free ($d\tilde{r}/d\tau\approx-\tilde{r}^{-1/2}$). In this limit the collapse also becomes independent of $\gamma$.

\section{Mapping From CMF to IMF}\label{sec:mapping}

In this section we discuss an algorithm for mapping an initial CMF to a simulated IMF. For that we carry out several Monte Carlo simulations, which calculate the time evolution of last crossing surfaces around a randomly chosen point in a collapsing medium. This means solving the stochastic differential equation of Eq. \ref{eq:delta_evol} for the case of a collapsing protostellar core.

\begin{figure}
\begin {center}
\includegraphics[width=\linewidth]{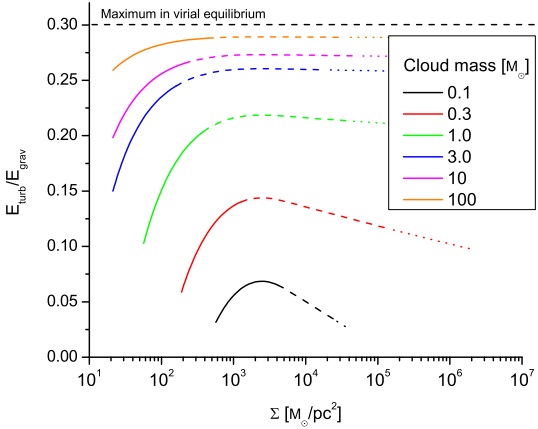}
\caption{Evolution of the ratio of turbulent to gravitational energy as a function of surface density in clouds during collapse ($\gamma\left(\Sigma\right)$ from Eq. \ref{eq:gamma} used as EOS). The solid, dashed and dotted lines show the evolution during the first 50\%, 90\% and the entirety of the collapse time (collapse is achieved when the cloud size reached $10^{-4}\,\pc$ which is roughly the size of a protostar). It is apparent that smaller clouds are mainly supported by thermal pressure and the relative importance of turbulence increases as the cloud collapses until $\gamma=4/3$ is reached (at $\Sigma\approx 2500\,\solarmass/\pc^2$ for this EOS) after which thermal energy grows faster than turbulent energy and starts dominating (see Eq. \ref{eq:mach_edge_evol}). For this plot $E_{\rm turb}\sim M\frac{v_{\rm t}^2}{2}$ and $E_{\rm grav}\sim M\frac{5 G M}{3 R}$. Virial equilibrium implies $c_s^2(1+\mach^2)=GM/R$ leading to $\frac{E_{\rm turb}}{E_{\rm grav}}\sim \frac{3 \mach^2}{10(1+\mach^2)}$ which sets 0.3 as the theoretical maximum.}
\label{fig:energyevol}
\end {center}
\end{figure}

\begin{figure}
\begin {center}
\includegraphics[width=\linewidth]{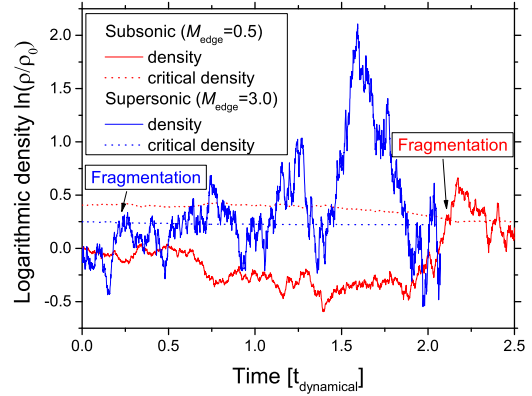}
\caption{Time evolution of the averaged density (smoothed on some sub-scale $\lambda$ around a specific random point within a cloud) and the critical density on the same scale (the density above which a region of this size becomes independently self gravitating). The curves follow a region whose size evolves with the parent cloud (it is a constant fraction of the parent cloud size). We consider both a supersonic (blue) and subsonic (red) cloud. The density field follows an essentially random walk. The first time it reaches the critical threshold, the subregion becomes self gravitating and starts to collapse on its own, thus fragmenting the cloud.}\label{fig:density_evol}
\end {center}
\end{figure}

In our simulation the cores start out internally homogeneous (this is a good approximation for the density and temperature below the last crossing scale of a full galaxy calculation) and start to collapse following Eq. \ref{eq:scale_evol}. As Fig. \ref{fig:energyevol} shows this leads to increased turbulence, which in turn leads to large density fluctuations (Eq. \ref{eq:S_def}). Through pumping turbulence, the collapse also modifies the critical density (Eq. \ref{eq:collapse_threshold_T}), combined with the aforementioned density fluctuations, this can lead to the formation of self gravitating subregions and thus the fragmentation of the parent cloud (see Fig. \ref{fig:fragment_fig}). Fig. \ref{fig:density_evol} shows the time evolution of the averaged and critical density on a specific scale for a subsonic and a supersonic cloud. The first time the density reaches the critical density on some scale, a self gravitating subregion appears, which is subsequently assumed to evolve independently from the parent cloud. This assumption is supported by the fact that the collapse timescale $t_0\sim (GM/\lambda^3)^{-1/2}\propto 1/\sqrt{\rho}$ and $\rho_{\rm crit}>\rho_0$ so smaller regions collapse faster, meaning that a small fragment can form a protostar much sooner than its parent could.

Based on these assumption our model follows the scheme:
\begin{enumerate}
	\item Initialize a cloud (e.g. density and temperature distribution).\label{itm:first}
	\item Evolve the density and temperature (assuming locally polytropic behavior) on all scales within the cloud until the first collapsing subregion appears (see Fig. \ref{fig:density_evol}).
	\item If there is a self gravitating subregion, evolve it forward starting again from step \ref{itm:first} using the parameters of the fragment at the moment of fragmentation as initial conditions.
\end{enumerate}
This scheme yields the so called \textit{collapse history} which contains the time evolution of the last crossing scale around a point. It is important to note that this model makes no assumptions about the relative position of the fragment within the parent cloud, thus what we calculate is the collapse history of a random point. By carrying out a large number of these simulations we can determine the statistical collapse history of a random Lagrangian point for a specific initial cloud. In other words we calculate the probability that a Lagrangian point/volume element inside the cloud ``ends up'' in a final fragment of some mass.

\begin{figure}
\begin {center}
\includegraphics[width=\linewidth]{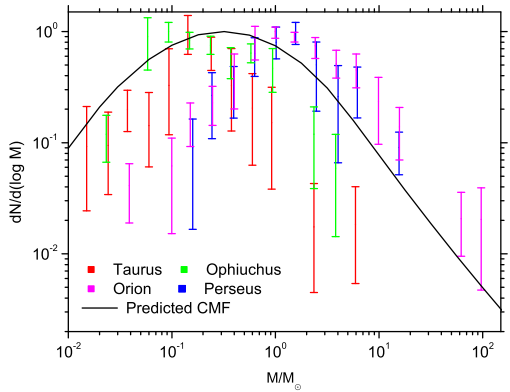}
\caption{Comparison between the CMF used in our calculations (the result of the excursion set model from \protect\citealt{core_IMF}) and a compilation of observed core mass functions from \protect\cite{Sadavoy_observed_CMF}. Since the exact scaling of the CMF is determined by the sonic mass, which depends on the parameters of the galactic disk, it was set in a way that the CMF turnover mass is between the observational limits. Effects of deviations from this default CMF are investigated in Sec. \ref{sec:CMF_deviate}.}
\label{fig:CMF_observ}
\end {center}
\end{figure}

The initial clouds represent the smallest self gravitating structures formed by fully developed turbulence in a galactic disk, which we consider to be equivalent to the observed protostellar cores. Their distribution has been calculated by \cite{core_IMF} using the same excursion set formalism, which naturally predicts their global parameters (see Fig. \ref{fig:CMF_observ}). By definition these clouds ``start out'' at the critical density so according to Eqs. \ref{eq:rho_crit_supersonic}-\ref{eq:rho_crit_subsonic} in the supersonic limit $M_{\rm core}\propto \lambda_{\rm core}^{p}$ (we took $p=2$ for the turbulent power index in our simulations) meaning a constant surface density $\Sigma$, and thus constant $\gamma(\Sigma)$ (see Sec. \ref{sec:eq_state}). Meanwhile in the subsonic limit $M_{\rm core}\propto \lambda_{\rm core}^{3-2/(2-\gamma)}$ which we can further approximate by taking the isothermal $\gamma=1$ case yielding $M_{\rm core}\propto \lambda_{\rm core}$. To get absolute scales let us assume virial equilibrium at cloud's scale which yields $c_s^2+v_t^2(R)\sim G M/R$. Now we can introduce the sonic scale $R_{\rm sonic}$, which correspond to the scale where $v_t^2\left(R_{\rm sonic}\right)=c_s^2$, and the sonic mass $M_{\rm sonic}$ which is the minimum self-gravitating mass contained in this subregion of size $R_{\rm sonic}$. These assumptions lead to the following mass-size relation for the initial cores:
\be
\label{eq:mass_size_sonic}
R(M) =\begin{cases} R_{\rm sonic}\frac{M}{M_{\rm sonic}}\,  & M<M_{\rm sonic} \\
										R_{\rm sonic}\sqrt{\frac{M}{M_{\rm sonic}}}  & M>M_{\rm sonic} \\
		  \end{cases}
\ee
By substituting in typical values for cores ($T=30\,\rm{K}$, $R\sim 0.1\,\pc$, see \citealt{MacLow_star_formation_ISM}) we get $M_{\rm sonic}\sim 3\,\solarmass$ for the sonic mass and
\be
\label{eq:mass_size}
R(M) =\begin{cases} 0.1\frac{M}{3 \solarmass}\,\pc  & M<3 \solarmass \\
										 0.1\sqrt{\frac{M}{3 \solarmass}}\,\pc  & M>3 \solarmass \\
		  \end{cases}
\ee
Note that the predicted size-mass relation agrees with that observed \citep{Larson_law, Bolatto_2008, Pineda_CMF_var}; we would obtain nearly identical results if we simply took the observed relation as our input.

Since the protostellar core in question has not yet started collapsing, the turbulent velocity at its edge must (initially) obey the turbulent power spectrum. Thus $v_t^2(R)\propto R$ for the supersonic and $v_t^2(R)\propto R^{2/3}$ (the Kolmogorov scaling) for the subsonic case. Using the mass-size relations of Eq. \ref{eq:mass_size_sonic} leads to the following fitting function
\be
\frac{\left(1+\mach^2_{\rm edge}\right)\mach^2_{\rm edge}}{1+\mach^{-1}_{\rm edge}}=\frac{M}{M_{\rm sonic}},
\ee
which exhibits scalings of $M\propto\mach^3$ for the subsonic and $M\propto\mach^4$ for the supersonic case respectively, and (coupled to the size-mass relation above) very closely reproduces the observed linewidth-size relations (\citealt{Larson_law, Bolatto_2008, clustering_Lada}).

\begin{figure}
\begin {center}
\includegraphics[width=\linewidth]{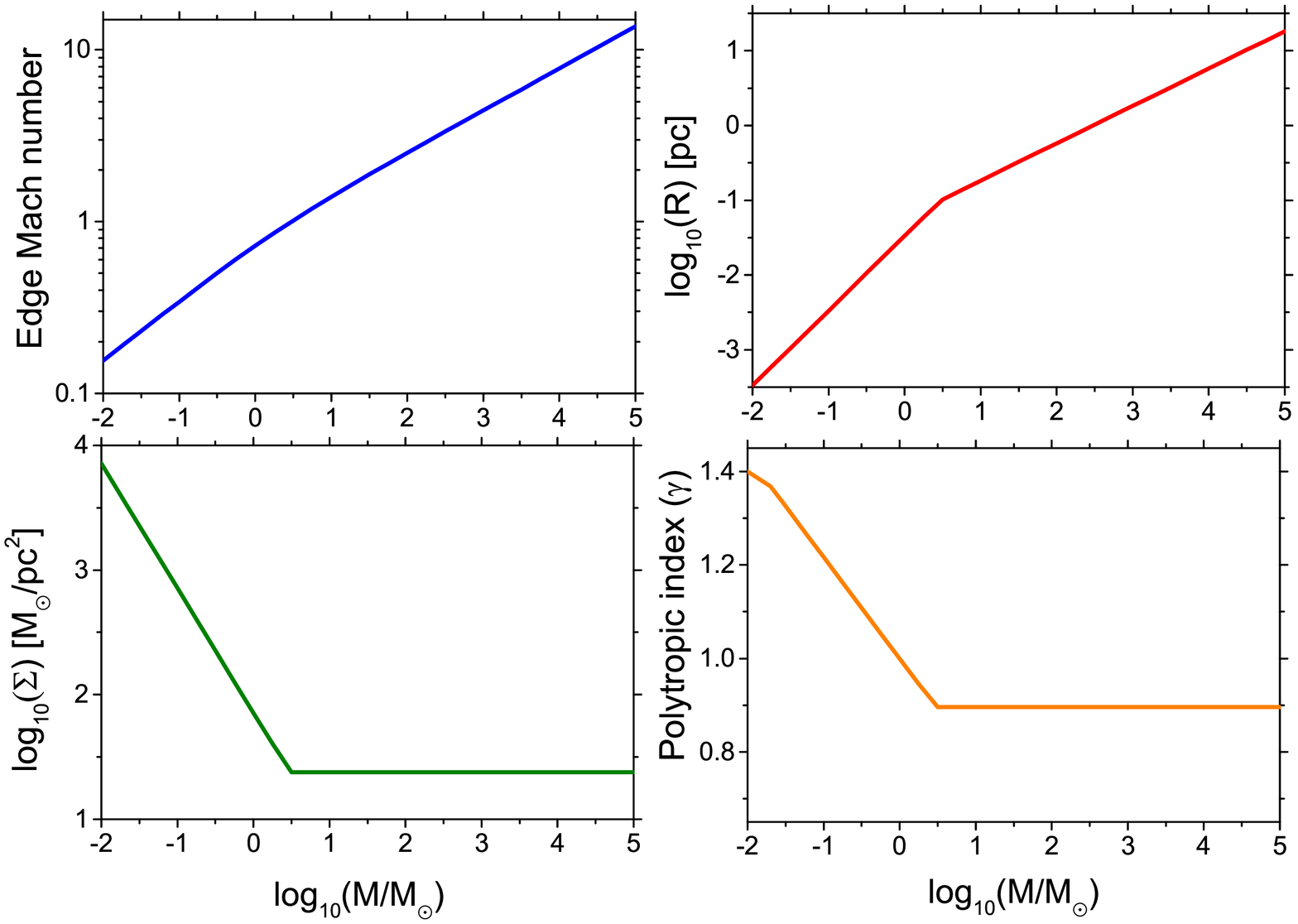}
\caption{Mass dependence of the initial parent core properties for the clouds on the observed CMF, used as the initial conditions for our calculation. We show the initial cloud scale or ``edge'' Mach number (top left), cloud radius $R$ (top right), cloud-averaged surface density $\Sigma$ (bottom left), and effective polytropic index $\gamma$ (bottom right) for protostellar cores before the collapse begins, each as a function of the initial core mass. These are calculated from the same excursion-set models from which the CMF in Fig.~\ref{fig:CMF_observ} is derived. But the mass-size relation we adopt agrees well with Larson's law for both small and large masses (\protect\citealt{Larson_law, Pineda_CMF_var, Bolatto_2008}) as does the Mach number-mass relation (or equivalently, the linewidth-size relation).}
\label{fig:mass_relations}
\end {center}
\end{figure}

\begin{figure}
\begin {center}
\includegraphics[width=\linewidth]{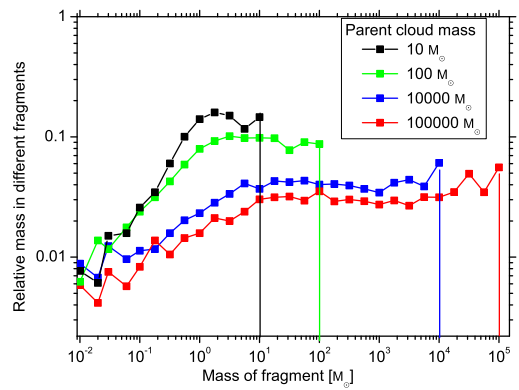}
\caption{Distribution of final (successfully collapsed to formally ``infinite'' density) fragments of different masses (total mass of fragments per logarithmic interval in fragment mass $d M/d\log{M_{\rm fragment}} = M\,dN/d\log{M_{\rm fragment}} = M^{2}\,d N/dM_{\rm fragment}$ so it is trivial to get $d N/dM_{\rm fragment}$ which is a more natural observable). We consider this for initial parent cores with different masses (and the surface density-dependent equation of state from Eq. \ref{eq:gamma}). Massive fragments can form (albeit rarely) without sub-fragmentation. In all cases where the parent is sufficiently large, there is a flat distribution ($d N/dM\propto M^{-2}$, approximately) at high fragment masses $\gtrsim M_{\odot}$, which is cut off at the mass of the parent cloud. This self-similar mass function owes to the fact that this is the ``scale free'' regime where turbulence and gravity dominate. The stiffer equation of state at higher densities, and sub-sonic nature of turbulence on small scales, suppress the number at low masses. Although only a small fraction of mass ends up in these fragments, this corresponds to a large number of individual stars. Also, a significant amount of mass ends up in substellar sized fragments which may either be destroyed by feedback mechanisms or form gas giants.}
\label{fig:volpdf}
\end {center}
\end{figure}

This means that an initial parent cloud can be described with only one physical parameter, which we chose to be its mass (see Fig. \ref{fig:mass_relations}). Using the aforementioned Monte Carlo algorithm it is possible to calculate $P_V(M_0,M)$ which is the probability that a randomly chosen initial Lagrangian point, within a parent core with initial mass $M_0$, ends up in a fragment of mass $M$ after collapse (see Fig. \ref{fig:volpdf}). Thus $P_V=0.1$ means that 10\% of the initial points thus 10\% of the total mass will end up in fragments of size $M$. The number of initial subregions containing $M$ mass is just $M_0/M$ so assuming the subregions are independent, the expected number of fragments becomes $P_V(M_0,M) M_0/M$. Thus, if the CMF is given by $n_{\rm core}(M)$ then the stellar IMF is
\be
\label{eq:imf}
n_{\rm stars}(M)=\int_{M}^{\infty}{n_{\rm core}(M')P_V(M',M) \frac{M'}{M} dM'}.
\ee
It should be noted that the CMF have significant uncertainties (\citealt{Pineda_CMF_var}); to account for that the effect of variations in the CMF are investigated in Sec. \ref{sec:CMF_deviate}.

It should be noted that Eq. \ref{eq:imf} neglects two important effects: geometry and feedback. Geometry becomes important as more fragments collapse to stars leaving behind "holes" in their parent cloud which hinders the formation of large scale substructures. This is related to the so called ``sphere packing problem'' that only a fraction of a sphere's volume (e.g. parent cloud) can be filled by non-overlapping spheres\footnote{Preliminary results from spatially resolved simulations suggest that these geometric effects cause only order of unity differences.}. Furthermore, Eq. \ref{eq:imf} assumes stars form independently and have no feedback on their parent cloud. This is not the case, especially if numerous small fragments form. We can imagine that when a protostar forms, it heats a region around it preventing that region from collapsing and forming protostars, with some mass $M_{\rm exc}$ which we call the \textit{exclusion mass}. We can crudely account for this effect by by taking the number of independent regions to be $M_0/M\rightarrow M_0/(M+M_{\rm exc})$. Essentially this ``excludes'' $M_{\rm exc}$ mass from further collapse each time a protostar forms.

What is a reasonable choice for the exclusion mass? \citet{Krumholz_stellar_mass_origin} argue that young, low-mass protostars accrete gas at a very high rate (leading to a luminosity $L\propto G\,M\,\dot{M}/R$ which grows rapidly in time) until they reach the mass required for deuterium burning, which leads to a characteristic luminosity and correspondingly, a characteristic mass of the surrounding median-density cloud which can be heated to the point where it is no longer gravitationally unstable. In their argument, depending on the background pressure, this produces an effective ``exclusion mass'' which varies between $10^{-2}-10^0\,\solarmass$. Based on this as a first approximation we will experiment with an exclusion mass of $\ord\left(0.01\,\solarmass\right)$. It should be noted that our intention with this crude assumption is not at all to give a full account of stellar feedback but to provide a simple correction mechanism for the overabundance of small mass fragments. In future work, we will explore a more self-consistent accounting for feedback in these calculations.

Another uncertainty is introduced by the fact that protostellar discs can fragment creating further brown dwarf sized objects. This combined with the sensitivity of the low mass end of the IMF to the equation of state of the gas and the crude approximation of feedback means that the model is highly uncertain in the very low mass region of the IMF.

\begin{figure}
\begin {center}
\includegraphics[width=\linewidth]{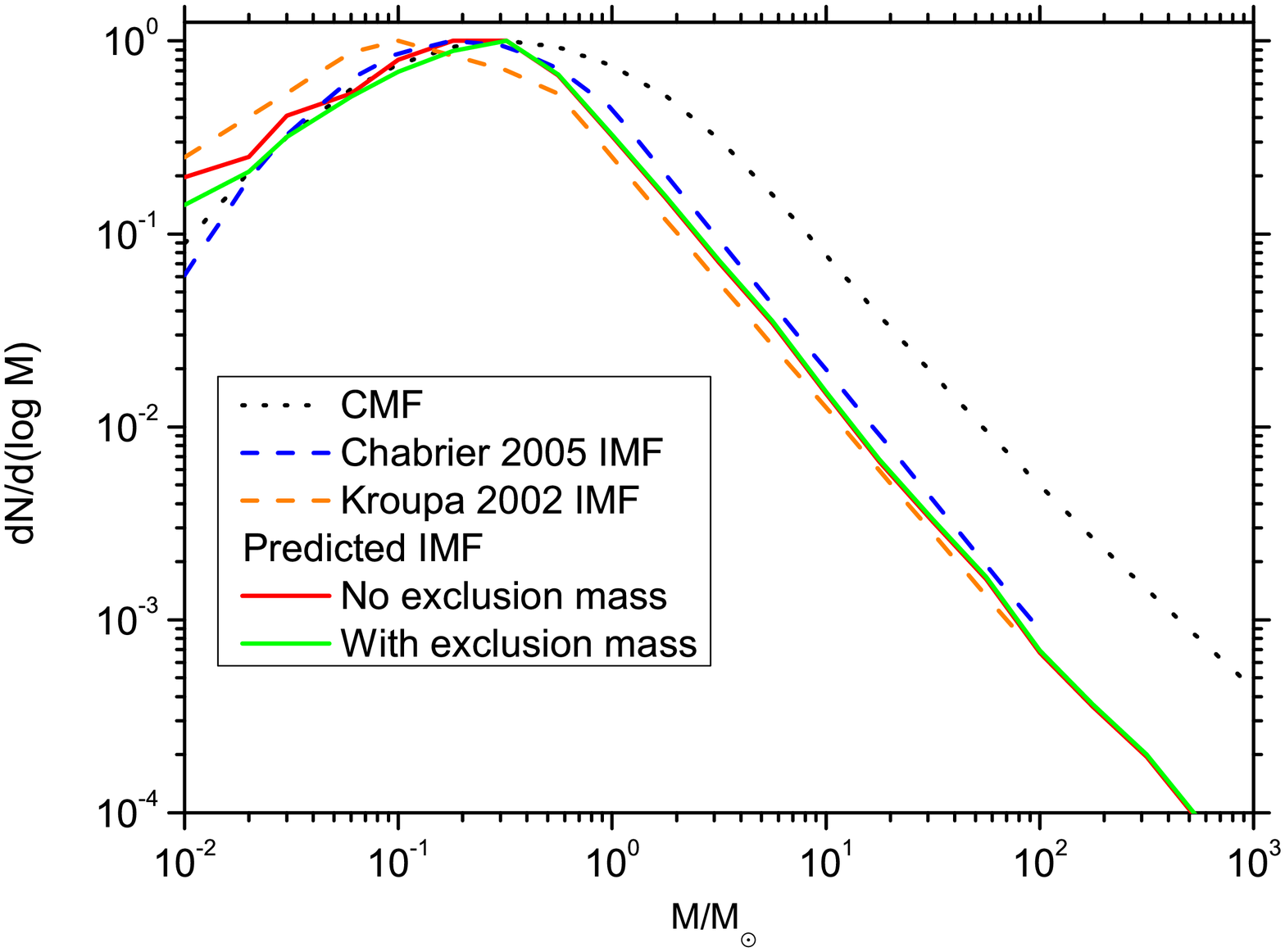}
\caption{Core mass function before and after final collapse compared with IMFs by \protect \cite{Kroupa_IMF} and \protect\cite{Chabrier_IMF}. Note that the absolute number (vertical normalization) is arbitrary, so we normalize each to the same peak value. After collapse/fragmentation, the high mass slope becomes slightly steeper, and the turnover point and cutoff mass move to lower masses. The model provides a near perfect fit to the observed IMF at the high mass end (the predicted slope of $2.32$ is well within the error of the nominal $2.35$). The calculations here use the surface density-dependent equation of state Eq.~\ref{eq:gamma}; this preserves the turnover at low masses; crudely the difference resembles a ``shift'' of the IMF peak by a factor of $\sim2-3$. However even in this case, there is some pile-up at small masses $<0.1\,M_{\sun}$, which may disagree with observations (depending on the preferred ``correct'' IMF); this can be mitigated by applying an appropriate exclusion mass (here we show the results for $M_{\rm exc}=0.02\,\solarmass$), which accounts for the protostars heating up their surroundings and preventing fragmentation.}
\label{fig:PDF_evol}
\end {center}
\end{figure}

We now consider the results of our calculation. Figure \ref{fig:PDF_evol} shows the core mass function before any collapse ($n_{\rm core}(M)$) and after collapse ($n_{\rm stars}(M)$). Compare this to the three qualitative properties of the IMF mentioned in Sec. \ref{sec:intro}. We find that it exhibits
\begin{enumerate}
	\item a power law scaling of of $\ord\left(M^{-2}\right)$ for high masses;
	\item turnover at $\ord\left(0.5\,\solarmass\right)$;
	\item close to lognormal dependence at low mass scales.
\end{enumerate}
In summary, it seems that this excursion set formalism can reproduce the main qualitative features of the IMF, and potentially provide an explanation for the universality of these properties. In the following subsections we consider these properties in more detail.

	\subsection{Dependence of the IMF on System Properties and Robustness of These Results}
	
Considering the the ubiquity of these IMF features in nature, and the number of assumptions in the model, it is critical to investigate the robustness of our results. The two primary parameters of our model are the initial CMF, which is dependent on the parameters of the original galactic disk for which the pre-collapse ``last crossing scale'' calculation was carried out, and the equation of state, which is highly uncertain.
	
	\subsubsection{Dependence on the Equation of State}\label{sec:robust:EOS}
	
\begin{figure}
\begin {center}
\includegraphics[width=\linewidth]{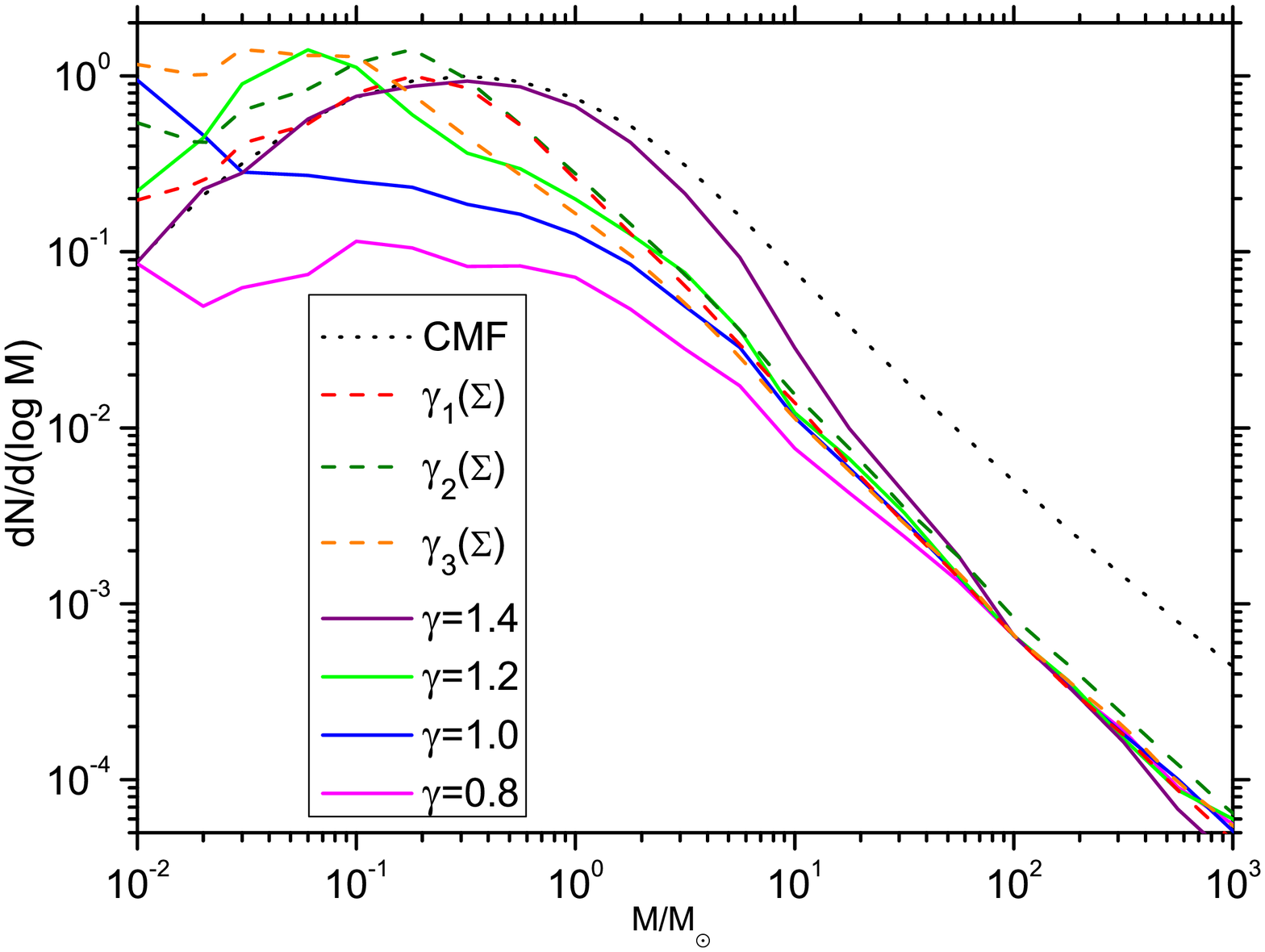}
\caption{Predicted IMF for different equations of state (constant polytropes, the original $\gamma\left(\Sigma\right)$ from Eq. \ref{eq:gamma}, and the shifted equations of state $\gamma_2\left(\Sigma\right)$ and $\gamma_3\left(\Sigma\right)$, where the upper surface density limit corresponding to $\gamma=1.4$ is set to $\Sigma=2\cdot 10^4\,\solarmass/\pc^2$ and $\Sigma=2\cdot 10^5\,\solarmass/\pc^2$ respectively, with no exclusion mass correction. The high mass end is insensitive to the choice of $\gamma$, as massive clouds are highly turbulent (see Eq. \ref{eq:rho_crit_supersonic}), leading to scale free fragmentation. We normalize the IMFs at $100\,M_{\sun}$ for ease of comparison. A ``soft'' EOS with $\gamma<4/3$ at all density scales would predict an excess (relative to observations) of fragmentation into brown dwarfs and sub-stellar objects ($M\lesssim 0.1\,M_{\sun}$). Some fragmentation can occur even with a ``stiff'' ($\gamma>4/3$) EOS, but only at very high masses where the turbulence is highly super-sonic. Lower $\gamma$ values lead to an increase in the number of small fragments, as there is less thermal pressure to resist fragmentation (see Eq. \ref{eq:collapse_threshold}).  Changing between the different functional forms of $\gamma\left(\Sigma\right)$ (which means increasing the upper density limit of the EOS) shifts the turnover point to lower masses and increases the number of small fragments as a higher surface density is required to reach high enough $\gamma$ values to resist further collapse.}
\label{fig:different_gamma_large}
\end {center}
\end{figure}

	First, we have repeated our calculations using different functional forms for the equation of state $\gamma$. Fig. \ref{fig:different_gamma_large} shows the resulting IMFs for constant $\gamma$ values (pure polytropes), for the original equation of state $\gamma_1\left(\Sigma\right)$ and for shifted equations of states ($\gamma_2\left(\Sigma\right)$ and $\gamma_3\left(\Sigma\right)$), where the upper surface density limit corresponding to $\gamma=1.4$ is set to $\Sigma=2\cdot 10^4\,\solarmass/\pc^2$ and $\Sigma=2\cdot 10^5\,\solarmass/\pc^2$ respectively (see Eq. \ref{eq:gamma} for original).
 
By analyzing the collapse histories, we have found that turbulent fragmentation occurs in a top-down cascade as large clouds fragment into clouds of smaller, but still comparable sizes (i.e.\ the largest scales tend to fragment first), which then undergo fragmentation again. Based on Fig. \ref{fig:different_gamma_large}, it is apparent that the high-mass power-law slope of the IMF is unaffected by the choice of $\gamma$, as all solutions tend to a power-law like slope which is slightly steeper than the original CMF slope, and is in good agreement with the observed Salpeter slope. That is because they are in the super-sonic regime (i.e. clouds have virial motions and/or initial turbulent motions which are firmly super-sonic); so the cloud dynamics and fragmentation are, to first order, dependent on turbulence and gravity, not on the thermal pressure of the gas, and the fragmentation cascade is inherently scale-free (as are both turbulence and gravity).

Note that our calculation predicts that ``final'' objects (which have successfully collapsed to infinitely high densities) can exist at high masses; successful collapse without fragmentation is rare, but not impossible. Because the cloud collapses in finite time, and the turbulent fluctuations are self-similar in the scale-free regime, the probability of avoiding a density fluctuation which would cause fragmentation is only power-law suppressed, not exponentially suppressed. Thus high-mass ``final'' cores can form. In fact our calculation predicts that the Salpeter slope continues to $\sim 10^{4}\solarmass$. If there is an actual ``maximum'' stellar mass -- i.e.\ if the actual stellar IMF cuts off at $\ord\left(100\,\solarmass\right)$, other factors besides pure turbulent fragmentation (e.g.\ fragmentation within the protostellar disk, or stellar stability at high masses, feedback from smaller stars, that form faster), must play a role. However whether such a cutoff exists is still uncertain.

\begin{figure}
\begin {center}
\includegraphics[width=\linewidth]{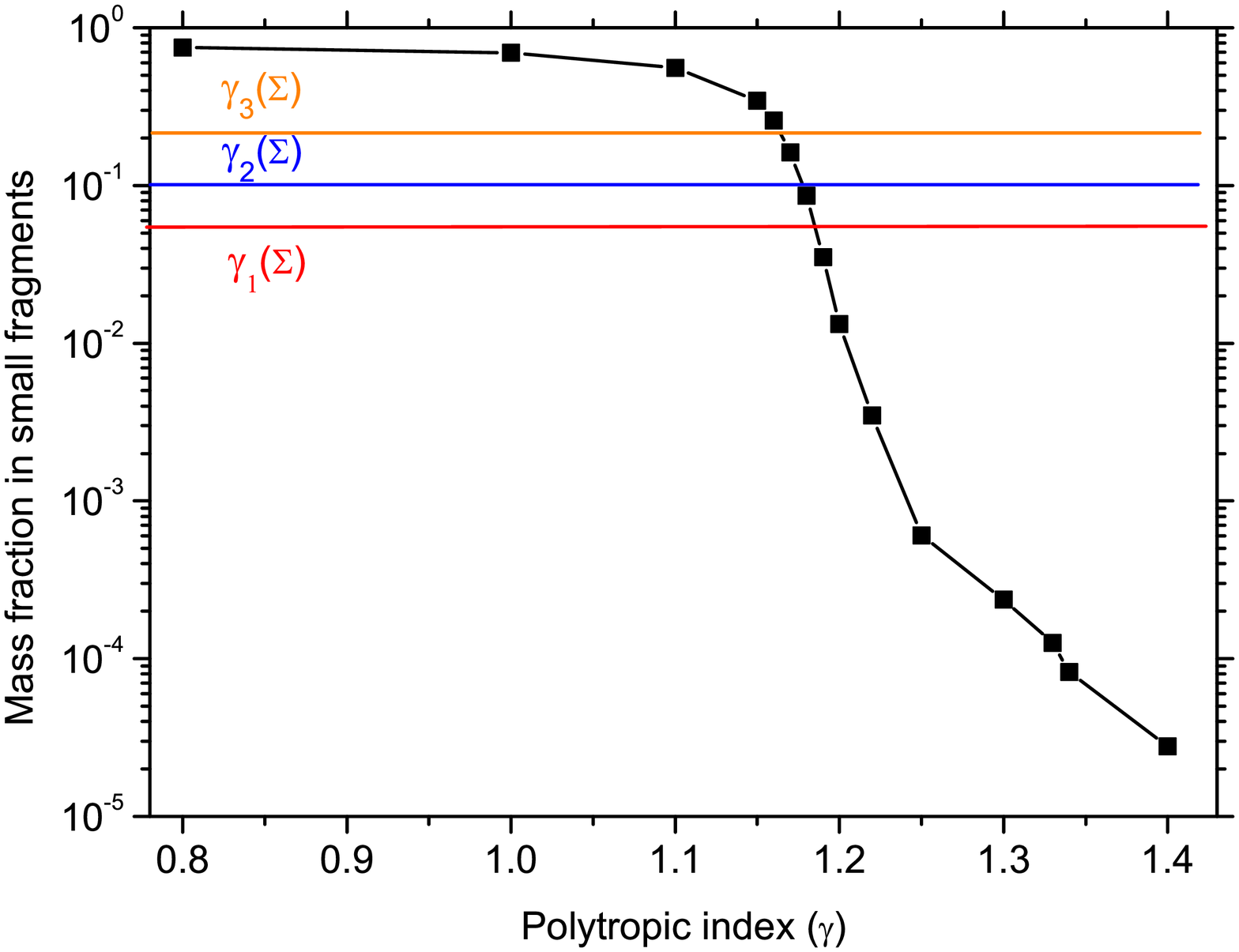}
\caption{Fraction of the total ``original CMF'' mass which ends up in sub-stellar ($M<0.01\,\solarmass$) fragments, for different equations of state. The single value lines correspond to our default (surface density-dependent) equation of state $\gamma\left(\Sigma\right)$ and the shifted $\gamma'\left(\Sigma\right)$; otherwise we assume a constant polytropic EOS and show the fraction as a function of that $\gamma$. For very soft (sub-isothermal) EOS values $\gamma<1$, the fragmentation cascades tend to proceed without limit, and most of the initial core mass ends up in arbitrarily small fragments! Higher $\gamma$ values allow the clouds to resist fragmentation, and above $\gamma=4/3$ small fragments basically vanish. No exclusion mass correction is applied here.}\label{fig:small_mass_fraction}
\end {center}
\end{figure}

Meanwhile, Fig. \ref{fig:different_gamma_large} also shows that the low-mass end of the IMF is heavily dependent on the equation of state. A stiff EOS ($\gamma>4/3$) basically freezes the CMF shape at solar and lower masses (no fragmentation occurs on small scales), while small values of $\gamma$ lead to increased fragmentation (Fig. \ref{fig:different_gamma_large}), which predict either no turnover in the IMF, or a turnover at much too-low masses. Note that in Fig. \ref{fig:different_gamma_large} it might at first appear that fragmentation is stronger in the $\gamma=1.0$ case than in the $\gamma=0.8$ case, however this is just an effect of the limited range and normalization of the plot, as there are actually a significant number of fragments which have smaller masses than $0.01\,\solarmass$ when $\gamma=0.8$. Fig. \ref{fig:small_mass_fraction} shows more clearly the fraction of the total mass ending up in substellar ($M<0.01\,\solarmass$) fragments, as a function of the EOS assumed\footnote{Our preliminary calculations with an explicitly 3D spatially dependent version of the model indicate that the substellar fraction is overestimated in Fig. \ref{fig:small_mass_fraction} because it is assumed that all mass ends up in bound structures while it is possible in reality for loose material to become unbound after fragmentation (see sphere packing considerations in Sec. \ref{sec:mapping}). The same discrepancy occurs in the cosmological Press-Schecter treatment, where it amounts to a factor of 2 at low masses. In case of our default EOS the difference is about a factor of 5.}. As expected, the formation of small fragments decreases monotonically with $\gamma$, and falls rapidly as we approach $\gamma=4/3$.

	\subsubsection{Dependence on the Core Mass Function}\label{sec:CMF_deviate}
		
The initial CMF used in our calculation is, itself, the prediction of turbulent fragmentation theory (it is the result of a similar excursion-set calculation of the ``last-crossing'' scales in a galactic disk; see \citealt{excursion_set_ism}). But the CMF could vary, or be different than predicted by this calculation owing to additional physics. We therefore next consider the IMF which results from different initial CMFs. 

To clearly isolate the most important dependencies and physics, it is actually much more instructive to adopt the following simple approximation of the CMF, rather than some more complicated functional form: 
\be
\frac{\mathrm{d} N}{\mathrm{d} \log{M}} \propto \begin{cases} M^{\alpha}  & M<M_{T} \\
																			M^{-\beta}  & M>M_T \\
											\end{cases},
\ee
where in our ``default'' CMF, $\alpha=1/2$ and $\beta=1.1$ are the approximate exponents of the low and high mass slopes, respectively, while $M_T=0.5\,\solarmass$ is the turnover mass. This allows us to systematically vary these three parameters and examine their impacts on the IMF. In each case, we will hold the equation of state $\gamma(\Sigma)$ fixed to our ``default'' value, and include no exclusion mass correction, so that the changes are purely a consequence of the CMF variation.

\begin{figure}
\begin {center}
\includegraphics[width=0.95 \linewidth]{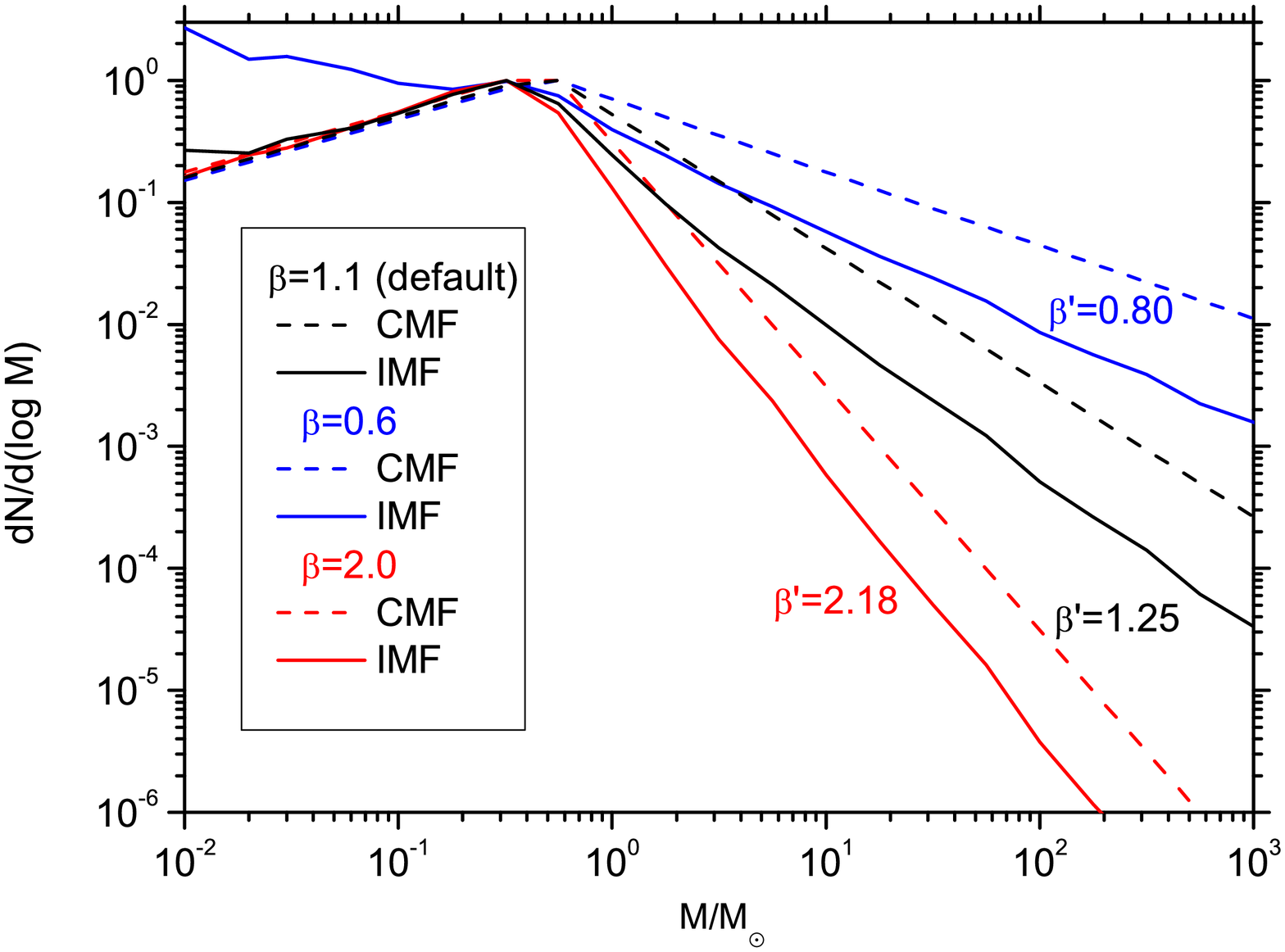}
\caption{Effects on the predicted IMF of having different slopes ($dN/d\log{M}\propto M^{-\beta}$) of the initial CMF. For each we keep all other parameters (e.g.\ $\gamma\left(\Sigma\right)$) fixed at their default values from Fig.~\ref{fig:PDF_evol}, and include no exclusion mass correction. We show the resulting IMF, with the final high-mass power-law ($\beta^{\prime}$) scaling. It is clear that fragmentation is close to scale-free as the IMFs produce high-mass power-law slopes close to the ``progenitor'' CMF slope, but steeper by a systematic $\Delta\beta\sim 0.2$. This systematic change can be understood as a consequence of time-dependent fragmentation at high masses; it also naturally explains the difference between the observed Salpeter slope of the IMF ($\sim 1.3$) and the predicted slope of the CMF from turbulent fragmentation models (closer to $\beta\approx 1.0-1.1$; see \protect\citealt{core_IMF, Hennebelle_theory}). Note that if the high-mass slope is sufficiently shallow ($\beta<1$), a pile-up at low masses results from fragmented large cores. However such shallow values are unphysical (they imply a divergent amount of mass in large cores).}
\label{fig:high_mass_CMF_var}
\end {center}
\end{figure}

In the turbulent framework we don't expect the high mass slope of the CMF ($\beta$) to vary as it is set by purely supersonic turbulence (see \cite{Hopkins_CMF_var}), however it is instructive to see whether the initial distribution (CMF) or the turbulent fragmentation sets the slope of the IMF. As Fig. \ref{fig:high_mass_CMF_var} shows fragmentation at the high mass end is close to scale free -- i.e.\ the slope of the IMF is always a power-law, which is systematically steeper than the CMF by a small, approximately fixed amount, independent of the actual initial high-mass slope of the CMF (or turnover mass, or low-mass CMF slope). The high-mass steepening is systematically $\Delta\beta\sim 0.2-0.25$. Let us consider now how much of a steepening would we expect. The IMF reflects the average rate at which final fragments collapse. The collapse time of a cloud is approximately $T_{\rm collapse} \sim t_{\rm dynamical} \sim 1/\sqrt{G M/R^3}$, which in the high-mass, supersonic limit ($R\propto \sqrt{M}$; see Eq.~\ref{eq:rho_crit_supersonic}) gives $T_{\rm collapse} \propto M^{1/4}$. So in the time for one high-mass core to collapse, multiple generations of low-mass cores can be spawned and collapse; to first approximation the ratio of the number of stars produced if we integrate over a fixed timescale (the collapse time of the large clouds) will be $n_{\rm stars}/\rm n_{\rm cores} \propto 1/T_{\rm collapse} \propto M^{-1/4}$, meaning $\Delta\beta=0.25$. 

\begin{figure}
\begin {center}
\includegraphics[width=0.95 \linewidth]{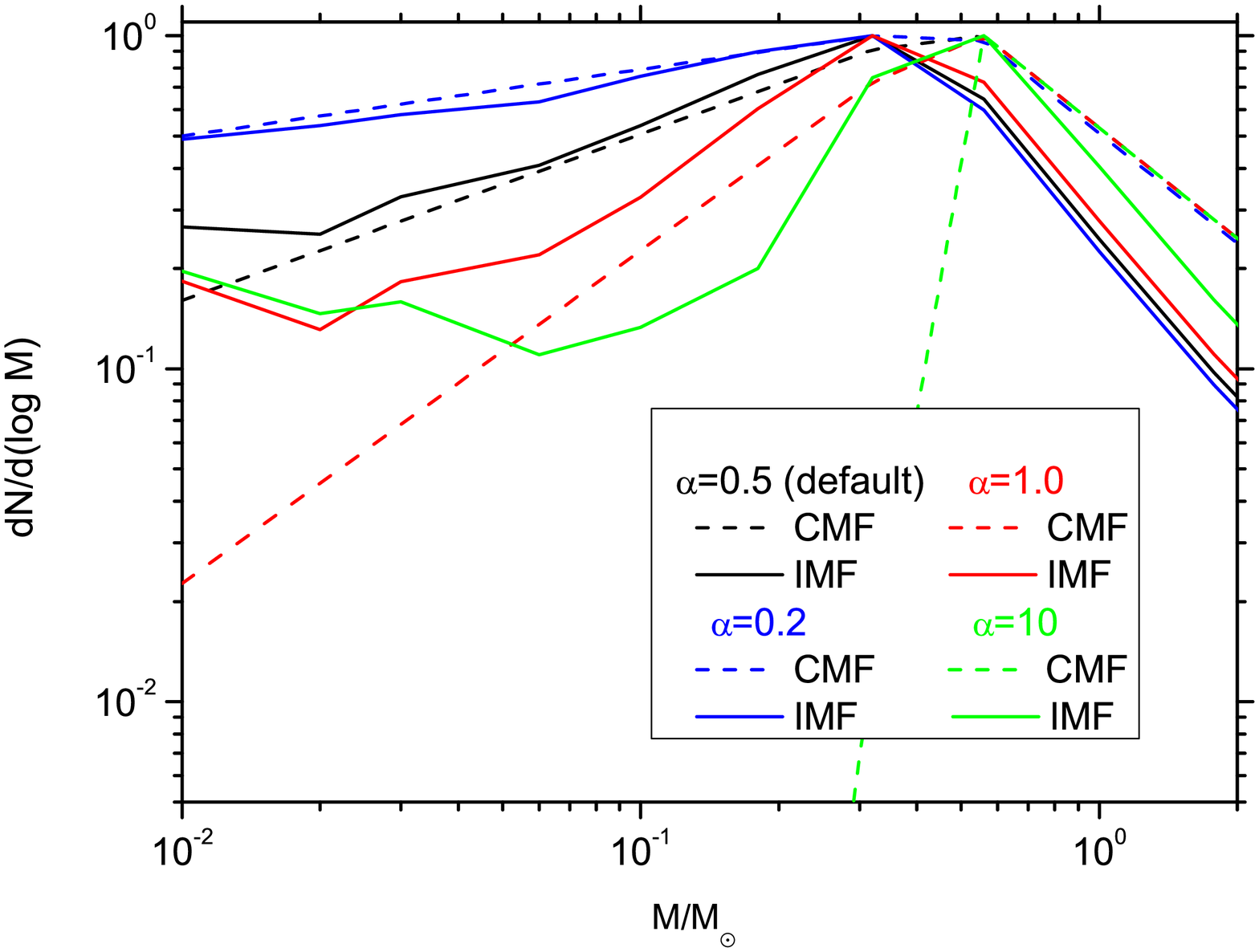}
\caption{Effects of different having slopes ($dN/d\log{M}\propto M^{+\alpha}$) at the low mass end of the initial CMF. As Fig.~\ref{fig:high_mass_CMF_var}, we keep all other parameters fixed. Since fragmentation is top-down, the low-mass CMF slope has no impact on the high-mass IMF. It is apparent that a significant fraction of the low-mass objects in the IMF are in fact fragments from much larger ``parent'' cores -- most clear when there are essentially no small cores to begin (the unphysical but instructive $\alpha=10$ case). However, for ``shallower'' initial CMF low-mass slopes, the IMF tends to trace the CMF, and the low mass stars are predominantly formed from low mass cores.}\label{fig:small_mass_CMF_var}
\end {center}
\end{figure}

The low mass end of the CMF is heavily dependent on galactic properties (see Fig. 2 of \cite{Hopkins_CMF_var}) so the value of $\alpha$ is far from fixed. However, small cores tend to collapse without further fragmentation so their effect on the IMF is just providing an initial population of small stars which is increased by the smaller fragments of high mass cores. This means that the low-mass end of the predicted IMF is sensitive to all changes in the CMF (Figs.  Fig. \ref{fig:high_mass_CMF_var}-\ref{fig:small_mass_CMF_var}). If we adopt an unphysical but instructive toy model where there are initially no low-mass cores, we see a sizable population of low-mass objects still appears in our final IMF. This is clear also from Fig.~\ref{fig:volpdf}; cores fragment into a very broad mass spectrum, and even high-mass cores can form very low-mass fragments. This is also evident if we adopt an initial CMF which has an (unphysically) shallow high-mass slope, such that there is an unlimited mass supply of very high-mass cores -- in turn there would be far too many small cores. It is also worth nothing that we appear to robustly predict that the approximate total number density ($dN/d\log{M}$) of objects with sub-stellar masses ($\sim 0.01-0.1\,\solarmass$) is never much less than $\sim 10\%$ that of objects with $\sim 0.1-1\,\solarmass$.

Finally, the turnover mass of the CMF ($M_T$) is proportional to the the sonic mass $M_{\rm sonic}\sim c_s^2 R_{\rm sonic}/G$ which is set by both galactic and local properties. This means that there could be some variation in the CMF turnover point (as noted by \cite{Hennebelle_theory, core_IMF}) which is in agreement with the observations (see Fig. \ref{fig:CMF_observ}). Interestingly, the position of the turnover point in the initial CMF only determines the point where the IMF starts to ``flatten''; however the details here are also dependent on the underlying physics (e.g.\ the equation of state). Nevertheless we can say that the turnover mass for reasonable parameters resides around $\ord\left(0.5 \,\solarmass\right)$.

\begin{figure}
\begin {center}
\includegraphics[width=0.95 \linewidth]{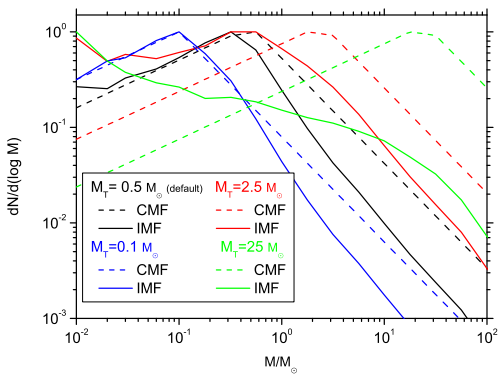}
\caption{Effects of moving the turnover mass ($M_T$) of the initial CMF. As Fig.~\ref{fig:high_mass_CMF_var}, we keep all other parameters fixed. The high-mass slope is unchanged by this choice, as in the turbulence-dominated regime the behavior becomes scale free (see Eq. \ref{eq:rho_crit_supersonic}). But the turnover point of the predicted IMF (or more accurately, where the resulting IMF becomes ``shallow'' and the total mass in stars converges, even if the IMF it does not completely turn over) clearly scales here with the turnover mass of the CMF.}\label{fig:turnover_CMF_var}
\end {center}
\end{figure}
	
\section{Conclusions}\label{sec:conclusions}

The aim of this paper was to provide a feasible candidate for the primary physical phenomena that determine the qualitative properties of the stellar initial mass function. This was achieved by expanding upon the excursion set formalism outlined in more detail by \cite{general_turbulent_fragment}, and applying it to follow the time-dependent collapse of protostellar cores into protostars. This improves on previous work done by \cite{Padoan_theory}, \cite{HC08} and \cite{core_IMF}, by following fragmentation down to stellar scales while taking into account the nonlinear time dependence and complicated equations of state (and their effects in making the density PDFs deviate dramatically from log-normal distributions). We found that this simple model reproduces the main qualitative features of the IMF, and it allows us to answer several critical unresolved questions in the theory of turbulent fragmentation.

The fact that both turbulence and gravity are scale free robustly predicts a CMF -- an instantaneous mass function of ``last-crossings'' -- with a high-mass slope $dN/dM\propto M^{-(2.0-2.1)}$ (see references above) -- this is the inevitable result of any scale-free, self-similar fragmentation process (basically, a slope of $-2$, which implies equal mass per log interval in mass, with a small logarithmic correction which depends on the properties of the medium but only very weakly). Time-dependent turbulent fragmentation slightly steepens this slope by a systematic $\Delta\beta=0.25$, creating a near-perfect fit with the canonical Salpeter slope of the observed IMF. The results are very robust to changes in both the initial conditions of the galactic disk, the equation of state of the gas, the presence of stellar feedback, the strength of the turbulence, and the form of the CMF. Thus we can say that the Salpeter slope is an inevitable consequence of turbulent fragmentation and is expected to be ``universal.''
	
Observed IMFs and CMFs have very similar shapes, and it appears as if the IMF is just a ``shifted'' version of the CMF. The simplest explanation would be that a constant fraction $\sim1/3$ of each core ends up in a single star. This is not the case in turbulent fragmentation. Rather, the apparent shift is the result of the nearly scale-free fragmentation in the high mass regime, and the flattening/turnover imprinted by the CMF and equation of state. We showed that, in fact, a high-mass core (which has initially no self-gravitating substructure) is typically expected to fragment into a broad range of masses, with comparable mass in fragments of all masses down to sub-solar masses. However, because this fragmentation produces a similar power-law slope for the IMF and CMF (see above), the result looks like a ``shift.'' We stress that the shift should not even be interpreted as an ``average fragment size'' -- that is actually much smaller (factor $<0.1$ of the original core size, for $\gg 10\,M_{\sun}$ cores). It is more accurate to say that sufficiently massive cores fragment into a spectrum of masses which resembles the IMF mass spectrum itself; since the convolution of a lognormal (or power-law) with another lognormal (or power-law) yields the same function, this produces the observed IMF shape. There is no one-to-one relation between cores and stars (far from it). 
	
It has been argued that purely isothermal turbulent fragmentation cannot produce the observed universal IMF, because collapsing clouds will inevitably become supersonically turbulent as gravitational energy pumps random motions, until fragmentation occurs. We confirm this is the case. Thus a CMF model based purely on isothermal turbulence -- or any simple polytrope -- is incomplete. However, that does not mean there could be no ``flattening'' of the IMF. Even for pure isothermal gas, the IMF still becomes more shallow than Salpeter around the ``sonic scale.'' This is related to what has been shown for the CMF: there is a characteristic scale in isothermal turbulence, the sonic scale, around which fragmentation becomes more or less ``easy.'' (It is only if one considers only thermal pressure, i.e.\ the Jeans length, that there is no characteristic scale). However, with isothermal gas, there is no true ``turnover'' in the IMF; moreover, most of the core mass ends up in very small (substellar) fragments.
	
We found that the turnover point in the initial CMF determines the point at which the IMF first ``flattens'' from the Salpeter slope. However, this does not necessarily amount to a full ``turnover'' in the IMF. Observationally, this ``flattening'' mass occurs at $\sim 0.5\,M_{\sun}$; for reasonable assumptions, we obtain a similar result. The CMF turnover point in turbulent fragmentation is robustly set by the ``sonic mass'' $M_{\rm sonic} \sim c_{s}^{2}\,R_{\rm sonic}/G$, the minimum self-gravitating mass at the sonic scale. Below this scale, the turbulence is sub-sonic, so large density fluctuations (in the parts of the medium which are not already self-gravitating) are not generated. As a result, we predict a ``flattening mass'' that scales as $\sim 0.5\,M_{\sun}\,(T_{\rm min}/30\,K)\,(\langle R_{\rm sonic}\rangle /0.1\,{\rm pc})$, where $T_{\rm min}$ is the minimum temperature reached by molecular cooling, and $\langle R_{\rm sonic}\rangle$ is the sonic length of the {\em pre-collapse} clouds -- i.e.\ the mean sonic length in the galactic disk (not a cloud-by-cloud quantity, since this changes as the cloud starts collapsing). As noted in \cite{Hopkins_CMF_var}, this predicts a very close to universal flattening mass within the Milky Way and nearby galaxies, but a lower flattening mass in extreme (high-Mach number) environments, where the sonic length is smaller, at the center of starburst galaxies and ellipticals. We will investigate this further in future work.

	The choice of equation of state, and effects of stellar feedback (crudely modeled here via an ``exclusion mass'' which is heated by each protostar) have some effect on the ``flattening mass,'' but a surprisingly weak one (shifting it by factors $\sim 2$, for a fixed CMF). However, they critically determine the behavior below this mass. In particular, whether the IMF actually ``turns over,'' or simply flattens, depends on these effects. If we assume any polytropic equation of state with $\gamma<4/3$, the IMF will still flatten, but will not turn over as observed (the IMF peak, in $dN/d\log{M}$, which is observed to be between $\sim 0.1-0.3\,M_{\sun}$, does not occur until $\ll 0.1\,M_{\sun}$). However, a surface-density dependent EOS, motivated by direct numerical calculations, is able to produce a reasonable turnover. This is because the characteristic surface density required for such a fragment to be self-gravitating is $\gtrsim 1\,{\rm g\,cm^{-2}}$ (higher if the fragment is embedded in an already-collapsing core, as we find is usually the case), so approaches the limit where it becomes optically thick to its own cooling radiation. If the this is indeed the relevant limit, we expect this mass to be weakly dependent on the minimum cooling temperature and the metallicity of the gas: requiring that a thermally pressure-supported cloud be self-gravitating, we predict this mass scales as $\sim 0.1\,M_{\sun}\,(T_{\rm min}/10\,K)^{2}\,(\kappa/\kappa_{\rm Milky Way})\sim 0.1\,M_{\sun}\,(T_{\rm min}/30\,K)^{2}\,(Z/Z_{\sun})$ -- this is weakly-varying in most systems, since $T_{\rm min}$ tends to decrease with metallicity (as low-temperature cooling is more efficient), while $\kappa$ increases. The presence of an ``exclusion mass'' further influences the details of the low-mass turnover, and may lead to a ``more universal'' behavior. We argue below that the key effects of feedback may be in preventing other effects we have ignored in our calculations.
	
	\subsection{Speculations}

We found that fragmentation usually occurs on a scale comparable to the parent cloud (because in turbulence, the power in density fluctuations is dominated by the large-scale fluctuations), which means that the fragmentation of collapsing cores can be accurately modeled as a top-down cascade. On average a large cloud loses equal amounts of mass to fragments per logarithmic interval in ``fragment mass'' (see Fig. \ref{fig:volpdf}), demonstrating the scale-free nature of the process. But even the largest, supersonic clouds have a nonzero chance of not fragmenting. This leads to an interesting prediction: turbulent fragmentation alone predicts that the Salpeter slope in a galactic disk continues to very high masses, $\sim 10^{4}\,M_{\sun}$. Whether such stars actually exist is still a matter of debate; however, it is commonly assumed that the most massive stars have masses $\sim 100-200\,\solarmass$. If this is the case, some other physics (e.g.\ fragmentation in proto-stellar disks, or stellar stability) must be the reason.
	
The fragmentation cascade predicted by the model can lead to the creation of substellar sized fragments, which could theoretically condense into gas giants. The amount of mass ending up in such fragments is heavily dependent  on the initial CMF and the equation of state. Nevertheless, it is important to note that these fragments would not be visible in numerical simulations (due to their resolution limits), but could lead to a large population of gas giant sized objects -- ``free-floating planets'' -- in the ISM. This, however, might not be the case if some physical process (e.g. stellar feedback) stops the cascade at smaller scales.
	
It should be noted that this model incorporates no real feedback physics, and does not take accretion by the protostars into account. Considering how well the results fit to the observed IMF, we tentatively conclude that those processes have negligible effect on the high-mass slope of the IMF. However, we believe stellar feedback could potentially solve the problem of the model predicting extremely massive ($\sim 10^{3}\,M_{\sun}$) stars. Since small objects collapse faster, there would be a significant number of realistic sized stars before a substructure of $10^{3}\,M_{\sun}$ could collapse. The more massive of these stars have a lifetime of several Myr which is comparable to the collapse time of the substructure. This means the cloud could be unbound by neighboring supernovae before it could collapse.

Meanwhile at the low-mass end, there is clearly a very strong effect from feedback, which we crudely modeled by way of either the ``effective equation of state'' or ``exclusion mass.'' However, even there, we do not necessarily expect feedback to strongly modify the ``top-down'' cascade we model. What may be more important, instead, is that feedback could prevent runaway accretion. Turbulent fragmentation naturally produces an IMF with the Salpeter slope and a turnover mass at the appropriate scale: subsequent ``competitive accretion'' would make the IMF more and more shallow, and turn sub-stellar fragments into brown dwarfs, leading to an excess population of such objects. The key role of feedback may therefore be to prevent such accretion -- i.e.\ ``shut down'' further accretion after the ``initial'' collapse (the part we model here). And in fact, this has been suggested in numerical simulations, where the ``initial'' IMF formed by turbulent fragmentation looks reasonable, but (without feedback) increasingly deviates from the observations as the system evolves (\citealt{Bate_lowmass_feedback, Offner_sim, Krumholz_stellar_mass_origin, Bate_sim}). 
	
\subsection{Future Work and Caveats}

Of course, although this model represents a qualitative improvement on the previous work in this area, further work is needed:

\begin{itemize}
	\item Many of the above points have been suggested by simulations (\citealt{Offner_sim, Krumholz_stellar_mass_origin, Federrath_sim_compare} etc.), however our analytical model allows us to follow an arbitrarily large range of scales (well beyond the resolution of numerical simulations). With our analytic model, we can also obtain statistically robust results, and easily explore a huge parameter space. Nevertheless it is necessary to test these results in full radiation hydrodynamics experiments.
	\item Due to its simplicity the model ignores several physical processes which could have a significant effect on star formation. An obvious omission is accretion, however the results do reproduce the observed IMF remarkably well, so the question is: does it not matter? An extension of the model which includes accretion (like done by \cite{Veltchev_2011}) could answer that question.
	
	\item Our model ignores magnetic fields, which may be an acceptable approximation in the high mass limit where the clouds are supersonic, but in the subsonic case ambipolar diffusion could be a serious factor in the collapse of clouds. It may, however, be possible to implement the most important effects of magnetic fields into the model by integrating it into the ``effective'' equation of state.
	
	\item The fragmentation cascade predict a large number of substellar fragments which could potentially collapse into gas giants. In future work, we will investigate in more detail the formation and evolution of such fragments, and compare their statistics to observational constraints.
	
	\item Another key observable is the spatial correlation function of star clusters and young stars. In future work, we will extend the models here to explore these observational constraints.

\end{itemize}

\acknowledgments
We thank Ralf Klessen and Mark Krumholz for inspirational conversations throughout the development of this work.

Support for PFH and DG was provided by the Gordon and Betty Moore Foundation through Grant \#776 to the Caltech Moore Center for Theoretical Cosmology and Physics, an Alfred P. Sloan Research Fellowship, NASA ATP Grant NNX14AH35G, and NSF Collaborative Research Grant \#1411920. Numerical calculations were run on the Caltech computer cluster ``Zwicky'' (NSF MRI award \#PHY-0960291) and allocation TG-AST130039 granted by the Extreme Science and Engineering Discovery Environment (XSEDE) supported by the NSF.

\bibliographystyle{apj}
\bibliography{bibliography}



\end{document}